\documentclass[conference]{IEEEtran}
\IEEEoverridecommandlockouts

\usepackage[ruled,vlined]{algorithm2e}
\usepackage{amsmath}
\usepackage{amssymb}
\usepackage{amsthm}
\usepackage{blindtext}
\usepackage[justification=centering]{caption}
\usepackage[noadjust]{cite}
\usepackage{epsf}
\usepackage{epsfig}
\usepackage{etoolbox}
\usepackage{float}
\usepackage{graphicx}
\usepackage{hhline}
\usepackage{hyperref}
\usepackage{latexsym}
\usepackage{multirow}
\usepackage{psfrag}
\usepackage[usenames,dvipsnames]{pstricks}
\usepackage{pst-plot}
\usepackage{stfloats}
\usepackage{subfigure}
\usepackage{tcolorbox}
\usepackage{balance}

\graphicspath{ {figures/} }

\input{Jerry.def}

\newtheorem{thm}{Theorem}

\newtheorem{cor}[thm]{Corollary}

\begin{document}
\title{
Bandwidth-Constrained Distributed Quickest Change Detection in Heterogeneous Sensor Networks: Anonymous vs Non-Anonymous Settings 
{\thanks{
This work was supported by the National Science and Technology Council of Taiwan under grants MOST 111-2221-E-A49 -069-MY3 and MOST 111-2811-E-A49-519-MY3.}}
}

\author{\vspace{-1.5cm}\\ Wen-Hsuan Li and Yu-Chih Huang\\
Institute of Communications Engineering\\
National Yang Ming Chiao Tung University,  Taiwan\\
E-mail:\{vincent, jerryhuang\}@nycu.edu.tw
}

\maketitle
\thispagestyle{empty}
\pagestyle{empty}

\begin{abstract}
The heterogeneous distributed quickest change detection (HetDQCD) problem with {\it 1-bit feedback} is studied, in which a fusion center monitors an abrupt change through a bunch of heterogeneous sensors via anonymous 1-bit feedbacks. Two fusion rules, one-shot and voting rules, are considered. We analyze the performance in terms of the worst-case expected detection delay and the average run length to false alarm for the two fusion rules. Our analysis unveils the mixed impact of involving more sensors into the decision and enables us to find near optimal choices of parameters in the two schemes. Notably, it is shown that, in contrast to the homogeneous setting, the first alarm rule may no longer lead to the best performance among one-shot schemes. The non-anonymous setting is then investigated where a novel  weighted voting rule is proposed that assigns different weights to votes from different types of sensors. Simulation results show that the proposed scheme is able to outperform all the above schemes and the mixture CUSUM scheme for the anonymous HetDQCD, hinting at the price of anonymity.


\end{abstract}




\section{Introduction} \label{Introduction}
Quickest change detection (QCD) investigates efficient detection of an abrupt change under a tolerant false alarm constraint. Such a classical problem in statistical analysis and signal processing has numerous applications \cite{ref:Lai08, ref:Lakhina04, ref:Nizam16}, and several fundamental results have been proposed from early works \cite{ref:Page54, ref:Lorden71, ref:Moustakides86}. In particular, Page in \cite{ref:Page54} proposed the well-known cumulative sum (CUSUM) test. For the criterion in \cite{ref:Lorden71}, Lorden \cite{ref:Lorden71} and Moustakides \cite{ref:Moustakides86} proved the asymptotic and exact optimality of the CUSUM test, respectively.

With the rise of Internet of Things (IoT) and distributed learning/computing, the requirement of efficient anomaly detection has become exceptionally crucial, and QCD has found its applications in such distributed systems \cite{Ref:Liyan21_survey}. For example, the smart grid is suffered from the risk of cyberattack, and in order to avoid the damage due to critical voltage variation, one can detect the line outage with low latency by applying QCD algorithm on certain small-signal power model \cite{ref:Chen16}. However, the task of detection is usually carried out by a fusion center through distributed sensors who feedback their observations/decisions via band-limited links. To overcome bandwidth constraints, Mei \cite{ref:Mei05} proposed a family of schemes where the sensors feedback the binary messages that whether the local CUSUM statistics reach a certain predefined threshold. When the fusion center applied consensus rule to combine feedback messages, Mei \cite{ref:Mei05} verified the asymptotic optimality of this stopping rule. However, the simulations in \cite{ref:Mei05} and \cite{ref:Tartakovsky08} showed that the consensus rule is not exactly optimal. In \cite{ref:Banerjee16}, Banerjee and Fellouris considered the same problem and further investigated the performance of the one-shot and voting rules with various number of alarms. The second order analysis in \cite{ref:Banerjee16} indicated that the first alarm is the best choice among one-shot schemes, while the majority vote rule is a reasonably good strategy among the voting rules.

Another challenge in  distributed QCD is the difference among the quality of information sent from the sensors. This discrepancy could originate from several sources such as transmission error/noise, sensing abilities, cyberattacks, etc. Such a problem with  Byzantine attacks being the source of discrepancy has been investigated in \cite{ref:Fellouris18} and \cite{ref:Huang21}.
The heterogeneous distributed QCD (HetDQCD) problem in \cite{ref:Sun20} modeled the heterogeneity by assuming each sensor's observations are drawn from a possibly non-identical but independent distribution. Sun {\it et al.} in \cite{ref:Sun20} showed that the mixture CUSUM procedure is an optimal anonymous rule when the observations are directly available at the fusion center.

Two questions naturally arise from this premise: {\it 1)  How to anonymously detect the change efficiently in HetDQCD under bandwidth constraints? 2) Does knowing the identities of the sensors (i.e., non-anonymous setting) help us achieve a better performance?} In this paper, we partially answer the above two questions for HetDQCD with {\it 1-bit feedback}. To address the first question, both the one-shot and voting rules in the anonymous HetDQCD are considered and analyzed.
Our first order analysis reveals appropriate choices of the ratios between thresholds of different sensors in both the one-shot and voting schemes. Furthermore, based on our second order analysis, the (asymptotically) optimal choice of the number of alarms we should wait in the one-shot scheme and that of the number of votes in the voting rule are derived. Our results indicate that among the one-shot schemes, the first alarm scheme may no longer be the best, which is in sharp contrast to the homogeneous setting. To address the second question, two schemes which exploit the knowledge of sensors' identities are introduced. The first scheme only accept alarms from the most informative sensors, while the second one further generalizes the idea to weight the votes according to the informativeness of the observations. Simulation results show that the proposed weighted voting rule (even the former simple special scheme in some cases) can outperform all the schemes considered for anonymous HetDQCD and the mixture CUSUM, hinting at the price of anonymity.


The remainder of this paper is organized as follows. Section \ref{ProblemFormulation} states the network model and problem formulation. Section \ref{NewFusionRule} proposes the new fusion rules for non-anonymous setting. The analysis results and some discussions are shown in Section \ref{Analysis}, which are verified via computer simulations in Section \ref{SimulationResults}. Finally, Section \ref{Conclusions} concludes the paper. 

\section{Background} \label{ProblemFormulation}
In this section, we introduce the network model, the local detection rule adopted at each sensor, and two fusion rules considered in this paper. The problem studied in this paper is to carefully analyze these schemes and provide efficient detection rule for HetDQCD with 1-bit feedback based on the analysis.

\subsection{Problem formulation}
\begin{figure}[t]
	\centering
	\includegraphics[scale=0.35]{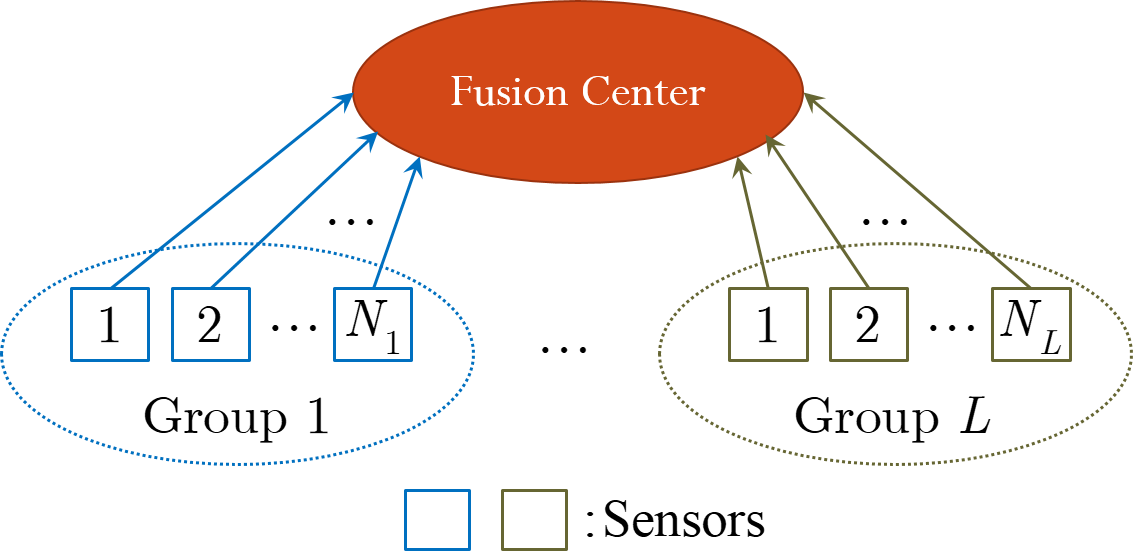}
	\caption{The heterogeneous wireless sensor network. \vspace{-0.2cm}\\}
	\label{network}
\end{figure}
We consider the heterogeneous sensor network as illustrated in Fig. \ref{network}, where $N$ sensors assist the fusion center in detecting an abrupt change occurring at time $\nu$. There are total $L$ heterogeneous types and thereby the $N$ sensors can be divided into $L$ groups, namely $\mathcal{G}_1, \ldots, \mathcal{G}_L$, according to their types. Let $N_l$ be the number of sensors of type $l\in[L]$ and we use the pair $(k,l)$ to denote the index of sensor $k\in[N_l]$ in group $l\in[L]$. The sensor $(k,l)$ will observe $X_{t}^{k,l}$ at time $t$ that is drawn independently according to
\begin{equation}
X_{t}^{k,l} \sim \left\{
                 \begin{array}{ll}
                   f_l, & \hbox{$t\leq\nu$,} \\
                   g_l, & \hbox{$t>\nu$,}
                 \end{array}
               \right.
\end{equation}
where $f_l$ and $g_l$ are the pre-changed and post-changed distributions, respectively, and they have the positive Kullback-Leibler divergence (KLD) $\mathcal{I}_{l}>0$. 
At time $t$, each sensor can communicate an 1-bit message with the fusion center noiselessly. A fusion rule $\rho$ is then adopted by the fusion center to form a decision according to the feedbacks. Specifically, let $\rho$ be an $\mathcal{F}_{t}$-stopping time, in which 
\begin{align}\mathcal{F}_{t}\equiv \sigma(X_{s}^{k,l};s \in [t],k \in [N_{l}],l \in [L]),
\end{align}
is the $\sigma$-algebra from all possible observations of all sensors up to time $t$. We denote by $P_{\nu}$ the underlying probability measure that the event happens at time $\nu$. In particular, if no event occurs, we set $\nu = \infty$ and denote by $P_{\infty}$ the underlying probability measure. Two metrics are defined to measure the performance of $\rho$, namely the average run length (ARL) to false alarm and the worst-case expected detection delay (EDD), as $\textrm{ARL}(\rho) = {E_{\infty}[\rho]}$, and
\begin{align}
    \textrm{EDD}(\rho) &=\sup_{\nu\geq 0}\mathrm{ess}\sup E_{\nu}[(\rho-\nu)^{+}|\mathcal{F}_{\nu}],
\end{align}
respectively, where $E_{\nu}$ is the expectation under $P_{\nu}$. Following Lorden's setting \cite{ref:Lorden71}, the problem of QCD is to minimize $\textrm{EDD}(\rho)$ subject to a constraint $\textrm{ARL}(\rho)>\gamma$.        

\subsection{Local CUSUM}
Similar to most of the work in the literature such as \cite{ref:Mei05, ref:Banerjee16, ref:Fellouris18}, each sensor first computes the CUSUM statistic locally. That is, let $Z_{t}^{k,l}$ and $W_{t}^{k,l}$ be the log-likelihood ratio (LLR) and the CUSUM statistic of the sensor $(k,l)$ at time $t$, respectively, each sensor recursively computes
\begin{align}
    W_{t}^{k,l} = \max{ \{0,W_{t-1}^{k,l} \}} + Z_{t}^{k,l} \textrm{, }\quad W_{0}^{k,l} \triangleq 0 \textrm{.}
\end{align}
To meet the bandwidth constraint, this CUSUM statistic cannot be losslessly sent to the fusion center and some process must be carried out. Due to the 1-bit feedback constraint, each sensor make the local decision whether $W^{k,l}_t>h_l$ at time $t$, where $h_l$ is the threshold for sensors in type $l\in[L]$.  


\subsection{Anonymous fusion rules} \label{Anonymous fusion rules}
At each time $t$, based on the feedback signals it has received, the fusion center makes a global decision about whether there is a change or not. Let $\mathbf{h} =[h_{1},...,h_{L}]^{T}$ be the threshold vector, two families of fusion rules are considered based on anonymous feedbacks \cite{ref:Banerjee16}:
\begin{itemize}
\item Anonymous $M$-th alarm: Each sensor sends a local alarm at the {\it first time} (hence, one-shot) its local CUSUM statistic exceeds a predefined threshold and a global decision is made when at least $M$ alarms are received. Mathematically, we have
\begin{align}
\rho^{(M)}( \mathbf{h} ) = \inf\{t: \rvert \{(k,l):\mathop{\max \limits_{t}}~W_{t}^{k,l}>h_l \} \rvert \geq M \} \textrm{.}    
\end{align}
\item Anonymous $M$ voting rule: Each sensor constantly sends the local decision and a global decision is made at the first time when there are at least $M$ alarms received simultaneously. Mathematically, we have
\begin{align}
    \rho_{M}( \mathbf{h} ) = \inf\{t: \rvert \{(k,l):W_{t}^{k,l}>h_l \} \rvert \geq M \}\textrm{.}
\end{align}
\end{itemize}

\section{Proposed Weighted Fusion rule} \label{NewFusionRule}

The two fusion rules in Section \ref{Anonymous fusion rules} treat every sensor equally, regardless of its type, and are thus suitable for the anonymous setting. Once the identities of the sensors are known to the fusion center, more potential fusion rules can be applied. A naive strategy is selecting the top informative sensors to avoid triggering the false alarm by other sensors. Given the predetermined set
$\mathcal{D}$ of selecting sensors, two fusion rules are proposed:

\begin{itemize}
    \item $M$-th alarm within $\mathcal{D}$:
    \begin{align}
&\tilde{\rho}^{(M)}( \mathbf{h},\mathcal{D}) = \nonumber \\
&\quad\inf\{t: \rvert \{(k,l)\in\mathcal{D}:\mathop{\max \limits_{t}}~W_{t}^{k,l}>h_l \} \rvert \geq M \} \textrm{.}
\end{align}
\item $M$ voting rule within $\mathcal{D}$:
\begin{align}
    \tilde{\rho}_{M}( \mathbf{h},\mathcal{D} ) = \inf\{t: \rvert \{(k,l)\in\mathcal{D}:W_{t}^{k,l}>h_l \} \rvert \geq M \}\textrm{.} \label{MVotingWithinD}
\end{align}
\end{itemize}

Notice that the sensor selection scheme can be applied even if the fusion center does not have the prior knowledge of distributions but the order of informative sensors. If furthermore, the fusion center obtains the information of KLD of each group, design of various weights for selecting sensor can further enhance the performance. Given $0 \leq \alpha_{k,l} \leq 1$ as the weight for the $k$th sensors in the $l$th group and let $\boldsymbol{\alpha} =[\alpha_{1,1},...,\alpha_{N_{L},L}]^{T}$, we can generalize the previous $M$ voting rule to the weighted version:
\begin{itemize}
\item Weighted $M$ voting rule:
\begin{align}
    \bar{\rho}_{M}( \mathbf{h},\boldsymbol{\alpha} ) =
    \inf\{t: \sum_{k,l} {\alpha_{k,l}\mathbf{1}{\{W_{t}^{k,l}>h_l\}} } \geq M \}
    \textrm{,}
\end{align}
\end{itemize}
in which $\mathbf{1} \{ \cdot \}$ is the indicator function. 

By setting $\alpha_{k,l} = \mathbf{1} \{ (k,l)\in\mathcal{D} \}$, the proposed weighted voting rule subsumes the selection scheme in \eqref{MVotingWithinD} as a special case.
As a heuristic, in our simulations, we propose setting the weights to be proportional to the KLDs, i.e., $\alpha_{k,l} \propto \mathcal{I}_{l}$ for $1 \leq l \leq L$.



\section{Analysis of Anonymous Fusion Rules} \label{Analysis} 
In this section, we investigate the performance of anonymous fusion rules. We first analyze the first order asymptotic performance of $\rho^{(M)}( \mathbf{h} )$ and $\rho_{M}(\mathbf{h})$, which guide us to choose appropriate thresholds $h_{l}$'s that lead to the best first order asymptotic scaling. Under the threshold setting, a second order analysis is then investigated, which enables us to choose the suitable $M$ in each family of anonymous fusion rules. Finally, using our analysis, the benefit of knowing the identities of sensors and the price of anonymity are discussed. The detailed analysis of non-anonymous fusion rules is left as future works.

Before proceeding to the analysis, we note that according to \cite[Lemma 3]{ref:Fellouris18}, any fusion rule $\rho$ based on the local CUSUM statistics (such as those above) admits a simpler expression of the EDD as
\begin{equation}
    \mathrm{EDD}(\rho)=E_0[\rho].
\end{equation}

\subsection{First Order Analysis}

Our first main result is the upper and lower bounds on the first order asymptotic of the expected detection delay as a function of the mean time to a false alarm for both $\rho^{(M)}(\mathbf{h})$ and $\rho_{M}(\mathbf{h})$.


\begin{thm} \label{asym_malarm}
Let $\mathbf{h} \mathnormal=[c_{1}h,\cdots,c_{L}h]^{T}$ with
$c_{1}/\mathcal{I}_{1} \leq c_{2}/\mathcal{I}_{2} \leq \dots \leq c_{L}/\mathcal{I}_{L}$, $c_{l}\in \mathbb{R}^{+}$ for all $1\leq l \leq L$. Suppose that $h = h_{\gamma,1}$ and $h = h_{\gamma,2}$ lead to $E_{\infty} [\rho^{(M)} (\mathbf{h})] = \gamma$ and $E_{\infty} [\rho_{M} (\mathbf{h})] = \gamma$, respectively.
Let $c^{(1)}\leq c^{(2)}\leq \dots \leq c^{(L)}$ be the rearrangement of $c_{1}, c_{2} \dots c_{L}$, and let $N^{(l)}$ and $\mathcal{I}^{(l)}$ represent the amount of the sensors in the group and KLD corresponding to $c^{(l)}$ respectively, $1\leq l \leq L$. Define $\lambda:[N] \rightarrow [L]$ by
\begin{align}
    \lambda (m) = l \textrm{~~if~~}\sum_{j=1}^{l-1}{N^{(j)}} \leq m \leq \sum_{j=1}^{l}{N^{(j)}} \textrm{.} \label{lambda_def}
\end{align}
Then as $\gamma \rightarrow \infty$,
\begin{align}
    &h_{\gamma,1} \sim \frac {\log \gamma} {c^{(\lambda (M))}}, \label{thres_malarm}\\
    &h_{\gamma,2} \leq \frac {\log \gamma} {\sum_{m=1}^{M}{c^{(\lambda (m))}}}(1+o(1)). \label{thres_mvote}
\end{align}
Moreover, we have
\begin{align}
    &E_{0}[\rho^{(M)}(\mathbf{h})] \in
    \left[\frac{c_{1}\log\gamma}{\mathcal{I}_{1}c^{(\lambda (M))}}(1+o(1)),
          \frac{c_{L}\log\gamma}{\mathcal{I}_{L}c^{(\lambda (M))}}(1+o(1))\right], \label{FARADD_malarm}\\
    &E_{0}[\rho_{M}(\mathbf{h})] \in
    \left[
    \frac{c_{1}\mathcal{I}_{\lambda(M)}\log\gamma(1+o(1))}{c_{\lambda(M)}\mathcal{I}_{1}\sum_{m=1}^{M}{\mathcal{I}^{(\lambda(m))}}},
    \frac{c_{L}\log\gamma(1+o(1))}{\mathcal{I}_{L}\sum_{m=1}^{M}{c^{(\lambda (m))}}}
    \right].
    \label{FARADD_mvote}
\end{align}
\end{thm}
\begin{IEEEproof}
    See Appendix.
\end{IEEEproof}

\vspace{0.1cm} We notice that in Theorem \ref{asym_malarm}, the ratio of thresholds between each group give a great effect in the upper and lower bound of detection delays in equations \eqref{FARADD_malarm} and \eqref{FARADD_mvote}, and one sub-optimal policy for choosing the thresholds is to minimize these upper bounds. This policy turns out to make the thresholds of each groups be proportional to the KLD of the pre-changed and post-changed distributions in this group. From now on, we will choose $c_{l} = \mathcal{I}_{l}$ for $1 \leq l \leq L$, and assume that $\mathcal{I}_{1} \leq \cdots \leq \mathcal{I}_{L}$ for simplifying notation. This induces the following corollary.

\begin{cor} \label{asym_malarm_cor}
Let $\mathbf{h} \mathnormal=[\mathcal{I}_{1}h,\cdots,\mathcal{I}_{L}h]^{T}$. Define $h_{\gamma,1}$ and $h_{\gamma,2}$ as in Theorem \ref{asym_malarm}. Then as $\gamma \rightarrow \infty$,
\begin{align}
    &h_{\gamma,1} \sim \frac {\log \gamma} {\mathcal{I}_{\lambda (M)}} \textrm{, } h_{\gamma,2} \sim \frac {\log \gamma} {\sum_{m=1}^{M}{\mathcal{I}_{\lambda (m)}}}. \label{thres_cor}
\end{align}

Furthermore,
\begin{align}
    &E_{0}[\rho^{(M)}(\mathbf{h})]
    \sim \frac{\log\gamma}{\mathcal{I}_{\lambda (M)}},
    \label{FARADD_malarm_cor}\\
    &E_{0}[\rho_{M}(\mathbf{h})]
    \sim \frac{\log\gamma}{\sum_{m=1}^{M}{\mathcal{I}_{\lambda (m)}}}. \label{FARADD_mvote_cor}
\end{align}
\end{cor}
\begin{IEEEproof}
Setting $c_{l} = \mathcal{I}_{l}$ for $1 \leq l \leq L$ completes the proof.
\end{IEEEproof}

\subsection{Second Order Analysis}
The first order analysis in Theorem \ref{asym_malarm} has enabled us to set the thresholds for obtaining the best asymptotic scaling as shown in Corollary \ref{asym_malarm_cor}. However, in order to compare the performance of different choices of $M$ under small to medium $\gamma$, we need a finer analysis. Here, we present the second order analysis which will later help us choose the best $M$.

\begin{thm} \label{2nd_order}
Let $\mathbf{h} \mathnormal=[\mathcal{I}_{1}h,\cdots,\mathcal{I}_{L}h]^{T}$. Assume that
\begin{align}
    \sigma_{l}^{2} \equiv E_{0}\left[ \left( Z^{k,l}_{t}-\mathcal{I}_{l} \right)^{2} \right] < \infty \textrm{ for all } k,l. \label{2nd_moment_assumption}
\end{align}
Then as $h \rightarrow \infty$,
\begin{align}
    &E_{0}[\rho^{(M)}(\mathbf{h})]= h + \xi _{M}\sqrt{h}(1+o(1)) \label{2nd_ADD_malarm}\\
    &E_{0}[\rho_{M}(\mathbf{h})] \leq
    h + \xi _{M}\sqrt{h}(1+o(1)), \label{2nd_ADD_mvote}
\end{align}
where $\xi_{M}$ is the expected value of $M$-th order statistics of independent (but not necessarily identical) Gaussian random variables $G_{k,l} \sim N(0,\sigma_{l}^{2} / \mathcal{I}_{l}^2)$ for $1 \leq k \leq N_{l}$ and $1 \leq l \leq L$.
\end{thm}
\begin{IEEEproof}
    See Appendix.
\end{IEEEproof}

\subsection{Discussions}
We discuss the appropriate choice of $M$ suggested by our analysis. With the asymptotic setting in {\eqref{thres_cor}}, \eqref{2nd_ADD_malarm} and \eqref{2nd_ADD_mvote} become
\begin{align}
    &E_{0}[\rho^{(M)}(\mathbf{h})]= \left( \frac{\log\gamma}{\mathcal{I}_{\lambda (M)}}
            + \xi_{M}\sqrt{\frac{\log\gamma}{\mathcal{I}_{\lambda (M)}}} \right)(1+o(1)), \label{2nd_ADD_malarm_final}\\
    &E_{0}[\rho_{M}(\mathbf{h})] \leq \nonumber\\
    &\left(
    \frac{\log\gamma}{\sum_{m=1}^{M}{\mathcal{I}_{\lambda (M)}}} + \xi_{M}\sqrt{\frac{\log\gamma}{\sum_{m=1}^{M}{\mathcal{I}_{\lambda (M)}}}} \right)(1+o(1)), \label{2nd_ADD_mvote_final}
\end{align}
respectively.

We first focus on the voting rule $\rho_M(\mathbf{h})$. For sufficiently large $\gamma$, the first order in  \eqref{2nd_ADD_mvote_final} will eventually dominate and one should set $M=N$ that minimizes the first term. On the other hand, for small to medium $\gamma$, due to the impact of $\xi_M$, one really has to solve for $M$ that minimizes \eqref{2nd_ADD_mvote_final}. For example, we can prove that $\xi_{M}$ is increasing in $M$ and is negative if $M<N/2$ and non-negative otherwise; thereby, the second order term in \eqref{2nd_ADD_malarm_final} and \eqref{2nd_ADD_mvote_final} will have negative impact on the detection delay when $M$ increases. The analysis of $\xi_M$ in the heterogeneous setting will allow us to quickly determine an appropriate $M$ for a finite $\gamma$ and is left for future work.

For the $M$-th alarm rule $\rho^{(M)}(\mathbf{h})$, things become very interesting even for large $\gamma$. Specifically, due to the piece-wise nature of $\lambda(\cdot)$ in \eqref{lambda_def}, any choice of $1+\sum_{j=1}^{l-1}N^{(j)} \leq M \leq \sum_{j=1}^{l}N^{(j)}$ for $1\leq l\leq L$ leads to the same first order term in  \eqref{2nd_ADD_malarm_final}. Within this region, one should pick
\begin{equation}\label{candidate_alarm}
    M= 1+\sum_{j=1}^{l-1}N^{(j)},
\end{equation}
that results in the smallest $\xi_M$. In particular, the best choice among all $M$ occurs at $M=N-N^{(L)}+1$. However, the best choice of $M$ for any fixed $\gamma$ again requires detailed analysis of $\xi_M$. We summarize our observations as follows.

\begin{cor} \label{thm_choosing_M}
For large $\gamma$, appropriate choices of $M$ for $\rho^{(M)}(\mathbf{h})$ and $\rho_M(\mathbf{h})$ according to  \eqref{2nd_ADD_malarm_final} and \eqref{2nd_ADD_mvote_final} are $M=N-N^{(L)}+1$ and $M=N$, respectively.
\end{cor}

In the sequel, we discuss the non-anonymous setting with the 1-bit bandwidth constraint, where the fusion center knows the identity of each sensor. The following result compares $\tilde{\rho}^{(1)}$ with $\rho^{(M)}$:
\begin{cor} \label{thm_group_selection}
When $N_L<N/2$, for large $\gamma$, the stopping rule $\tilde{\rho}^{(1)}(\mathbf{h},\mathcal{G}_L)$ that accepts the first alarm from the most informative group $\mathcal{G}_L$ achieves a better tradeoff than $\rho^{(M)}(\mathbf{h})$ for any $M$.
\end{cor}
\begin{IEEEproof}
Following the suggestion in Theorem \ref{thm_choosing_M}, the minimal detection delay for $M$-th alarm is achieved by setting $M=N-N_{L}+1$, which implies that as $\gamma$ large enough,
\begin{align}
    E_{0}[\rho^{(M)}(\mathbf{h})] \geq \left( \frac{\log\gamma}{\mathcal{I}_{L}} + \xi_{N-N_{L}+1}\sqrt{\frac{\log\gamma}{\mathcal{I}_{L}}} \right)(1+o(1)) \nonumber\\
    \overset{(a)}{>} \frac{\log\gamma}{\mathcal{I}_{L}}(1+o(1)) = E_{0}[\tilde{\rho}^{(1)}(\mathbf{h},\mathcal{G}_L)] \textrm{ when } N_{L}< N/2 \textrm{,} \label{compare_malarm}
\end{align}
where (a) holds since $\xi_{M}>0$ for $M>N/2$.
\end{IEEEproof}


For $\tilde{\rho}_M$ and $\rho_M$, we note that precise comparison still requires computation of $\xi_M$. However, since $\xi_M$ is an increasing function, it is likely that $\tilde{\rho}_M$ focusing on several most informative groups would outperform $\rho_M$ for any $M$ which blindly cooperate among all the sensors in the network.


\section{Simulation Results}\label{SimulationResults}

\begin{figure}[t]
\centering
   \includegraphics[width=3.2in]{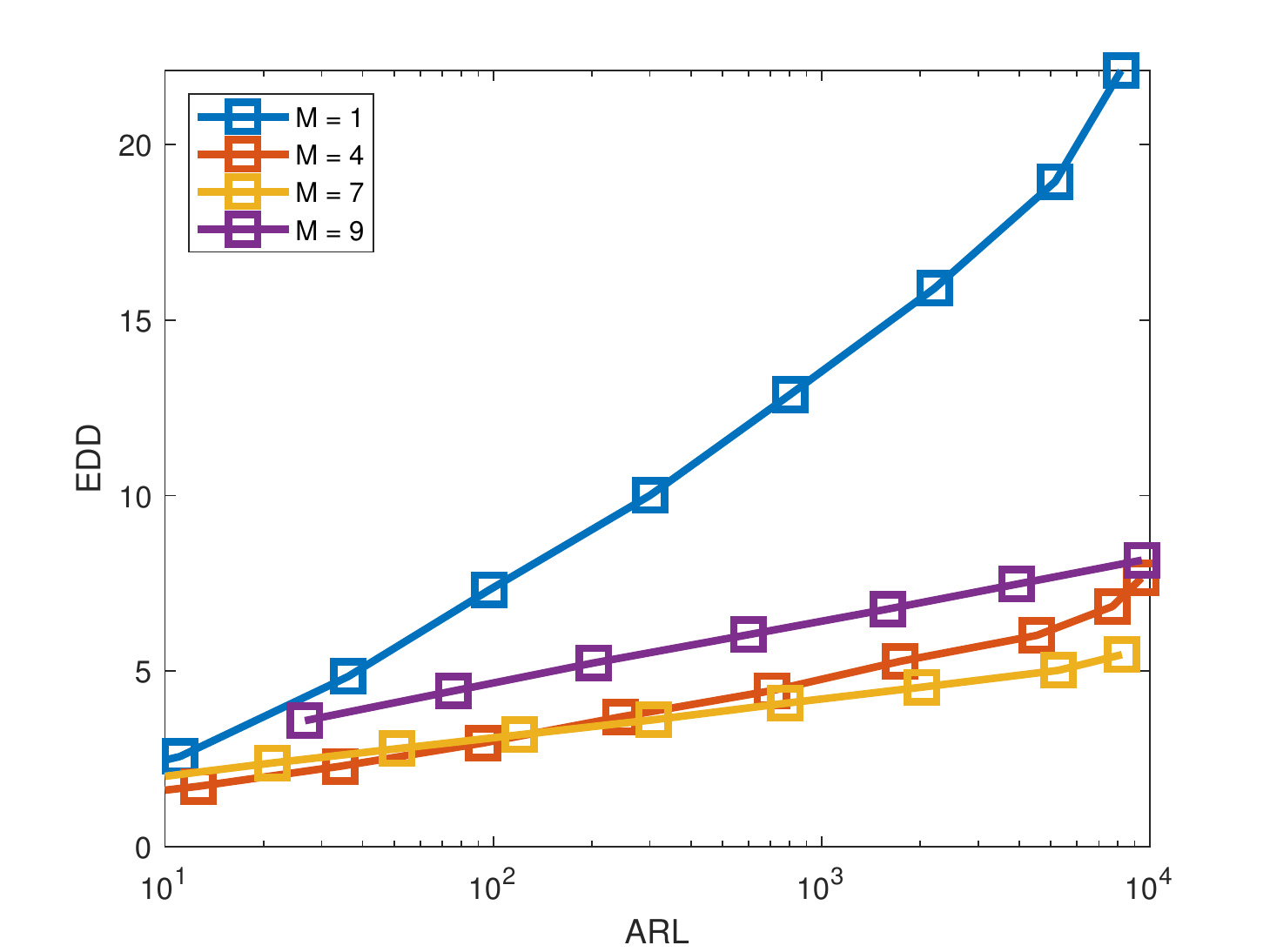}
   \caption{ARL versus EDD for $\rho^{(M)}(\mathbf{h})$. \vspace{-0.2cm}\\}
   \label{fig_m_alarm}
\end{figure}
\begin{figure}
\centering
   \includegraphics[width=3.2in]{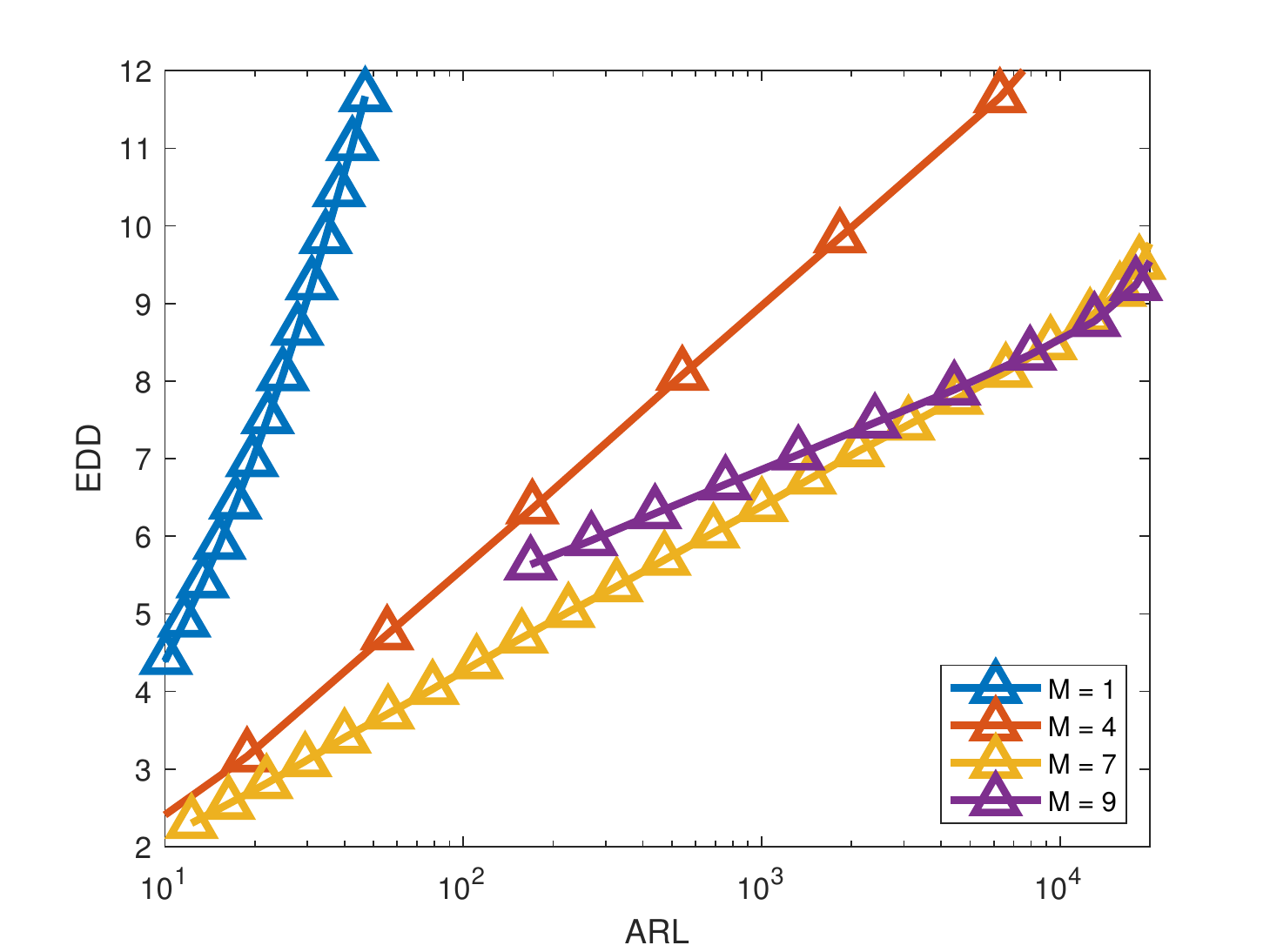}
   \caption{ARL versus EDD for $\rho_{M}(\mathbf{h})$. \vspace{-0.2cm}\\}
   \label{fig_m_vote}
\end{figure}

\begin{figure}[ht]
	\centering
	\includegraphics[width=3.2in]{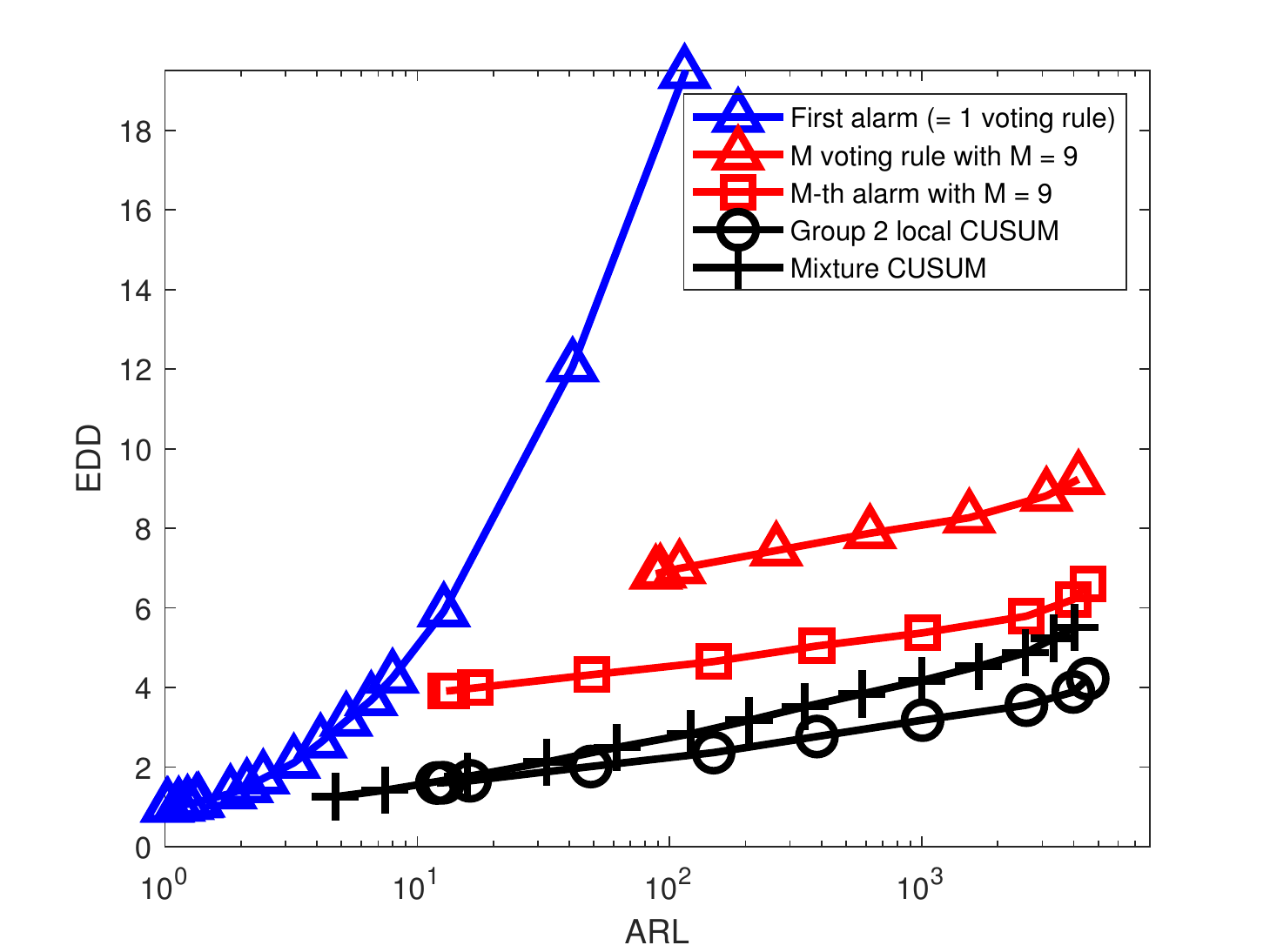}
	\caption{Comparison of $\rho^{(M)}(\mathbf{h})$, $\rho_{M}(\mathbf{h})$, and $\tilde{\rho}^{(1)}(\mathbf{h},\mathcal{G}_L)$.  \vspace{-0.2cm}\\}
	\label{fig_group_selection}
\end{figure}

\begin{figure}[ht]
	\centering
	\includegraphics[width=3.2in]{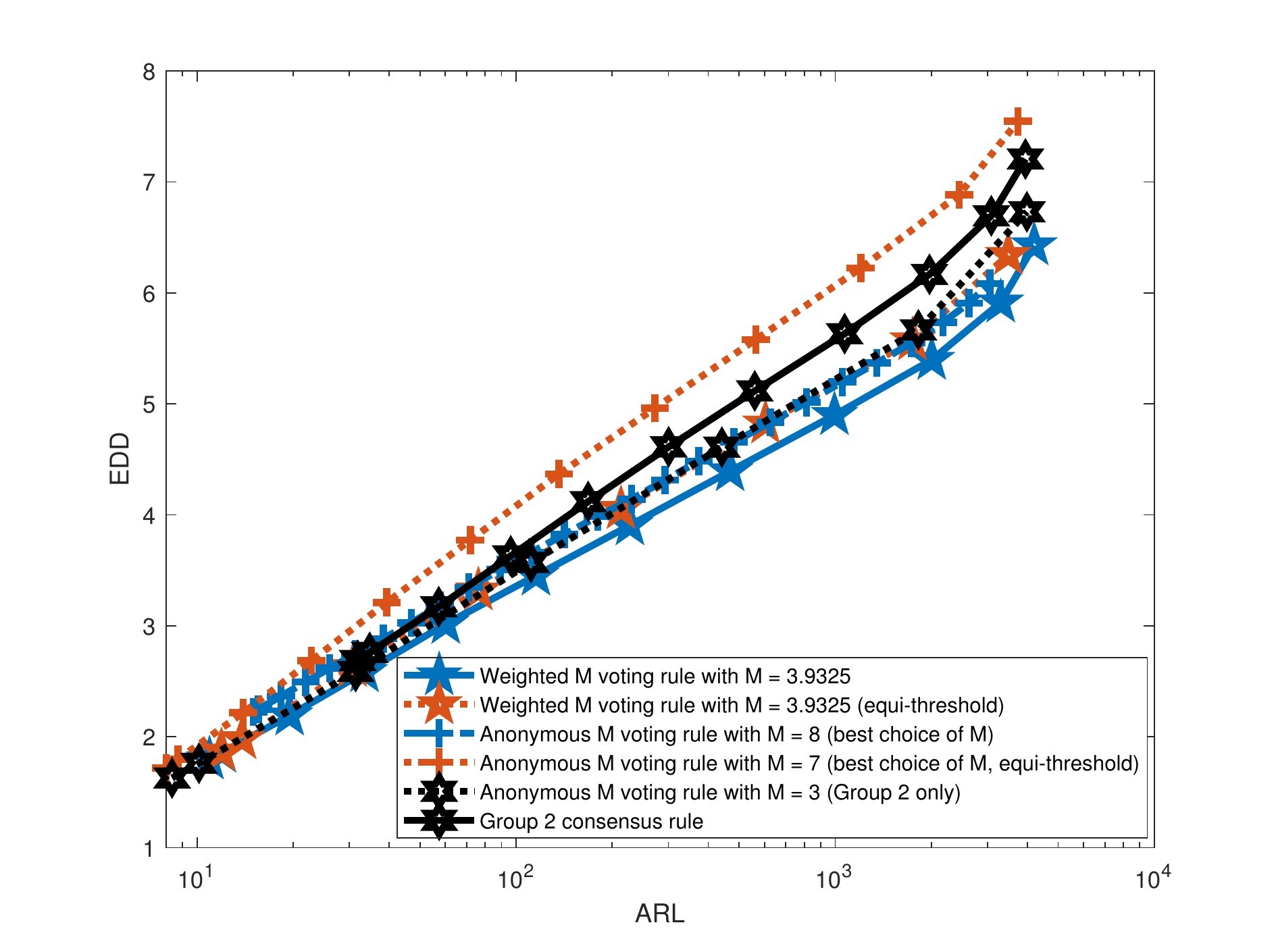}
	\caption{Comparison of weighted and anonymous $M$ voting rules. \vspace{-0.2cm}\\}
	\label{fig_opt_weight_voting}
\end{figure}

We first examine the discussion in Corollary \ref{thm_choosing_M}. Here we set $L=3$ and $N_{l}=3$ for each $l\in\{1,2, 3\}$. 
We first consider the $M$-th alarm rule $\rho^{(M)}(\mathbf{h})$ with $M\in\{1,4,7\}$ satisfying \eqref{candidate_alarm} and $M=9$ involving all the sensors. In this simulation, the pre-changed distributions remain standard normal for all sensors and the post-changed distributions of group $l$ are set to be $g_{l} \sim N(2l/3,l)$. 
As shown in Fig. \ref{fig_m_alarm}, when $\gamma$ is large enough, as predicted in Corollary \ref{thm_choosing_M}, the best two curves correspond to $M = 4$ and $7$ and $M=7$ leads to the best performance, indicating a major departure from the homogeneous setting where it was shown in \cite{ref:Fellouris18} that the first alarm is the best. Moreover, involving all the sensors (i.e., $M=9$) is also not a good choice. We then study the performance of the $M$ voting rules $\rho_M(\mathbf{h})$ with the pre-changed distributions being the standard normal and the post-changed distributions $g_{l} \sim N(l/3,1)$ for $l\in\{1,2,3\}$. We show in Fig. \ref{fig_m_vote} that when $\gamma$ is sufficiently large, $M=9$ is the best. However, when $\gamma$ is not large enough, $M=9$ is not always optimal, exhibiting the impact from $\xi_{M}$ in \eqref{2nd_ADD_mvote_final}.

The comparison between anonymous and non-anonymous settings is then investigated in Fig. \ref{fig_group_selection}. In this simulation, there are two groups consisting of $8$ and $1$ sensors, respectively. The pre-change distributions are again the standard normal distribution, while the post-changed distributions for group $1$ and group $2$ are set to be $g_{1} \sim N(0.
35,1)$ and $g_{2} \sim N(2,4)$, respectively. In this figure, the performance of $\rho^{(1)}(\mathbf{h})$, $\rho^{(9)}(\mathbf{h})$, and $\rho_{9}(\mathbf{h})$ are plotted. We note that we have tested $\rho^{(M)}(\mathbf{h})$ and $\rho_{M}(\mathbf{h})$ for all the choices of $M$ and $M=9$ is the best for both the families in this setting. In addition, the performance of the mixture CUSUM procedure is also provided. We note that this procedure has been shown to be optimal for the anonymous setting when the observations are directly available at the fusion center \cite{ref:Sun20}. Moreover, the performance of $\tilde{\rho}^{(1)}(\mathbf{h},\mathcal{G}_2)$ (equivalent to $\tilde{\rho}_1(\mathbf{h},\mathcal{G}_2)$ in this case as $N_2=1$) is plotted to reveal the benefit of non-anonymity. It is demonstrated that with the knowledge of sensors' identities, by simply ignoring the sensors in $\mathcal{G}_1$ and focusing solely on the sensor in $\mathcal{G}_2$, $\tilde{\rho}^{(1)}(\mathbf{h},\mathcal{G}_2)$ outperforms all the anonymous schemes considered, including the mixture CUSUM. This hints at the price of anonymity.

To further examine the performance enhancement of weighted voting and threshold setting, we demonstrate the simulation as shown in Fig \ref{fig_opt_weight_voting}. There are two groups with $6$ and $4$ sensors respectively. The pre-changed distribution still remains standard normal, and the post-changed distribution of group $1$ and group $2$ are $g_{l} \sim N(0.55,1)$ and $g_{2} \sim N(1,1)$ respectively. Following Section \ref{NewFusionRule}, the weights for weighted voting rule are set to be proportional to the KLDs, and the largest one is set to be unity. We can see that both the threshold and the weight settings improve the performance. In particular, the proposed weight give a better performance than equal weight (i.e. anonymous voting) or binary weight (i.e. group selection), which shows the potential of cooperation.

\section{Conclusions} \label{Conclusions}
In this paper, we investigated the HetDQCD problem with 1-bit feedback. For the anonymous setting, we rigorously analyzed two families of stopping rules, the $M$-th alarm rule and the $M$ voting rule. Using our analysis, an appropriate choice of $M$ for each scheme was suggested. The non-anonymous setting had also been discussed, where it was shown that focusing on the most informative sensors could outperform strategies taking in all the sensors. A weighted voting rule based on KLDs was proposed and simulation results were provided to demonstrate that it can outperform all the schemes based on anonymous reports considered in this paper, including the mixture CUSUM. Potential future works include a thorough investigation of the proposed $\tilde{\rho}^{(M)}(\mathbf{h},\mathcal{D})$ and $\tilde{\rho}_M(\mathbf{h},\mathcal{D})$ for any $\mathcal{D}$ in the non-anonymous setting, the optimal weight design for $\bar{\rho}_{M}( \mathbf{h},\boldsymbol{\alpha} )$, design of quantizers for HetDQCD with $b$-bit bandwidth constraint for $b>1$, and taking into account the presence of compromised sensors.

\clearpage
\balance
\bibliographystyle{IEEEtran}
\bibliography{reference}

\clearpage
\appendix \label{Appendix}
The following properties of local CUSUM shown \cite{ref:Fellouris18} will be useful for our proofs: 
\begin{align}
    P_{\infty} (W_{t}^{k,l} \geq h_{l}) \leq e^{-h_{l}},\label{FAR}\\
    E_{\infty} [\rho^{k,l}(h_{l})] \sim \mathbf{\Theta} (1)e^{h_{l}},\label{asym_FAR}\\
    P_{0} \left(\rho^{k,l}(h_{l}) \sim  \frac {h_{l}} {\mathcal{I}_{l} }\right) = 1,\label{ADD}
\end{align}
where $\rho^{k,l}(h_{l})$ is defined in \eqref{eq:def_stoptimelocal}.

\subsection{Proof of Theorem~\ref{asym_malarm}}

The proof of the theorem is done by combining the following important two lemmas, whose proofs are shown after this subsection.

\begin{lemma} \label{thm_FAR_bin}
Let $\mathbf{h} \mathnormal=[c_{1}h,\cdots,c_{L}h]^{T}$, $c_{l}\in \mathbb{R}^{+}$ for all $1\leq l \leq L$. Define $\lambda:[N] \rightarrow [L]$ by
\begin{align}
    \lambda (m) = l \textrm{~~if~~}\sum_{j=1}^{l-1}{N^{(j)}} \leq m \leq \sum_{j=1}^{l}{N^{(j)}} \textrm{.}
\end{align} 
Then, as $h \rightarrow \infty$,
\begin{align}
&E_{\infty} [\rho^{(M)} (\mathbf{h})] \sim \mathbf{\Theta}(1) e^{c^{(\lambda (M))}h}\textrm{,}\label{FAR_malarm}\\
&E_{\infty} [\rho_{M} (\mathbf{h})] \geq \mathbf{\Theta}(1) e^{(c^{(\lambda (1))}+...+c^{(\lambda (M))})h}\textrm{,} \label{FAR_mvote}
\end{align}
in which $c^{(1)}\leq c^{(2)}\leq \dots \leq c^{(L)}$ is the rearrangement of $c_{1}, c_{2} \dots c_{L}$.
\end{lemma}

\begin{lemma} \label{bd_ADD}
Let $\mathbf{h} \mathnormal=[c_{1}h,\cdots,c_{L}h]^{T}$ with
$c_{1}/\mathcal{I}_{1} \leq c_{2}/\mathcal{I}_{2} \leq \dots \leq c_{L}/\mathcal{I}_{L}$, $c_{l}\in \mathbb{R}^{+}$ for all $1\leq l \leq L$. Then, as $h \rightarrow \infty$, both $E_{0}[\rho^{(M)}(\mathbf{h})]$ and $E_{0}[\rho_{M}(\mathbf{h})]$ are asymptotically lower and upper bounded by $c_{1}h/\mathcal{I}_{1}$ and $c_{L}h/\mathcal{I}_{L}$, respectively.
\end{lemma}

\begin{IEEEproof} [Proof of Theorem~\ref{asym_malarm}]
Equations \eqref{thres_malarm} and \eqref{thres_mvote} follow directly from \eqref{FAR_malarm} and \eqref{FAR_mvote}, respectively. Plugging the thresholds of \eqref{thres_malarm} into the lower and upper bound in Lemma \ref{bd_ADD} results in the upper and lower limits of \eqref{FARADD_malarm}, respectively. Similarly, the upper bound of equation \eqref{FARADD_mvote} holds by plugging the thresholds in \eqref{thres_mvote} into the upper bound in Lemma \ref{bd_ADD}.

It remains to prove the lower bound of equation \eqref{FARADD_mvote}. We first focus on the case when $M = N$. The detection delay in this case is lower bounded by the performance of the well-known centralized CUSUM \cite{ref:Lorden71}, i.e.,
\begin{align}\label{eq_pf1}
E_{0}[\rho_{M}(\mathbf{h})] \geq
    \frac{\log\gamma}{\sum_{l=1}^{L}{N_{l}\mathcal{I}_{l}}} (1+o(1)).
\end{align}
Since $c_{l}/\mathcal{I}_{l} \geq c_{1}/\mathcal{I}_{1}$ for $l \geq 1$, \eqref{eq_pf1} implies the lower bound in \eqref{FARADD_mvote}. We go on to prove the case that $M<N$. Under the same stopping time rule, we compare the original network with its sub-network, in which the sub-network has exactly $M$ sensors choosing from the sensors corresponding to $M$ smallest thresholds $c^{(1)}h$ to $c^{(\lambda (M))}h$ . Denote  $h_{\gamma,3}$ be the threshold for the stopping time of this sub-network under the false alarm rate $\gamma$. Then, from the upper bound in Lemma \ref{bd_ADD} and applying \eqref{eq_pf1} to the sub-network, we have
\begin{align}
h_{\gamma,3} \geq
\frac{\log\gamma}{\sum_{m=1}^{M}{\mathcal{I}^{(\lambda (m))}}}
\cdot \frac{\mathcal{I}_{\lambda (M)}}{c_{\lambda (M)}}(1+o(1)). \label{hgammref:Fellouris18}
\end{align}
Since the original network contains the sub-network, we have $h_{\gamma,2} \geq h_{\gamma,3}$, and the lower bound in Lemma \ref{bd_ADD} implies
\begin{align}
E_{0}[\rho_{M}(\mathbf{h})] &\geq \frac{c_{1}}{\mathcal{I}_{1}} h_{\gamma,3} \nonumber\\
&\geq \frac{\log\gamma}{\sum_{m=1}^{M}{\mathcal{I}^{(\lambda(m))}}}
\cdot \frac{\mathcal{I}_{\lambda (M)}}{c_{\lambda (M)}}\cdot  \frac{c_{1}}{\mathcal{I}_{1}}(1+o(1)).
\end{align}
\end{IEEEproof}

\subsection{Proof of Lemma~\ref{thm_FAR_bin}}
\begin{IEEEproof}
From \eqref{asym_FAR}, the stopping time $\rho^{k,l}(h_{l})$ of each sensor are asymptotically exponential. Hence, \eqref{FAR_malarm} holds because \eqref{asym_FAR} and the independence of $\rho^{k,l}(h_{l})$ over every sensor. To show \eqref{FAR_mvote}, we first focus on the event that triggers the voting rule. We denote by $\Omega_M = \{ \pi_{j}^{M}: 1\leq j\leq \binom{N}{M} \}$ the collection of all subsets $\pi_{j}^{M}$ of ${(k,l)}$ with exactly $M$ 
pairs. Again, by the independence of stopping time among sensors, applying \eqref{FAR} shows that the false alarm probability of $M$ voting rule is bounded by
\begin{align}
    P_{\infty} (\rho_{M}(\mathbf{h}) < z ) &\leq 
    \sum_{t=1}^{z}{\sum_{\pi_{j}^{M} \in \Omega_M}{ \prod_{(k,l) \in \pi_{j}^{M}}{ P_{\infty} \left( W_{t}^{k,l} \geq h_{l} \right)}}}\nonumber\\
    &\leq \sum_{t=1}^{z}{\binom{N}{M}\max \limits_{\pi_{j}^{M}} \prod_{(k,l) \in \pi_{j}^{M}}{ P_{\infty} \left( W_{z}^{k,l} \geq c_{l}h \right)}}\nonumber\\
    &\leq z\binom{N}{M} \cdot e^{-(c^{(1)}+...+c^{(M)})h} \equiv z\gamma,
\end{align}
which leads to
\begin{align}
    E_{\infty} [\rho_{M}(\mathbf{h})] &= \int_{0}^{\infty}{P_{\infty} \left( \rho_{M}(\mathbf{h}) \geq z  \right)dz}\nonumber\\
    &\geq \int_{0}^{1/\gamma}{P_{\infty} \left( \rho_{M}(\mathbf{h}) \geq z \right)dz}\nonumber\\
    &\geq \int_{0}^{1/\gamma}{(1-z\gamma)dz} \geq \frac{1}{\gamma}.
\end{align}
\end{IEEEproof}

\subsection{Proof of Lemma~\ref{bd_ADD}}
\begin{IEEEproof}
From the definitions of stopping time for $M$-th alarm and $M$ voting rule, it is clear that
\begin{align}
    \rho^{(1)}(\mathbf{h})\leq \rho^{(M)}(\mathbf{h})
    \leq \rho_{M}(\mathbf{h})\leq \rho_{N}(\mathbf{h}).\label{inequ_stoppingtime}
\end{align} 
Therefore it suffices to show that
\begin{align}
    E_{0}[\rho^{(1)}(\mathbf{h})]\geq c_{1}h/\mathcal{I}_{1}\label{LADD_malarm}\\
    E_{0}[\rho_{M}(\mathbf{h})]\leq c_{L}h/\mathcal{I}_{L}\label{GADD_mvote}
\end{align}
We first prove \eqref{LADD_malarm}. For $1 \leq l \leq L$, denote by $\sigma_{l}(c_{l}h)=\min \limits_{k}\rho^{k,l}(c_{l}h)$. It follows that,
\begin{align}
    \rho^{(1)}( \mathbf{h} ) = \min \limits_{(k,l)}\rho^{k,l}(c_{l}h)=\min \limits_{l}\sigma_{l} (c_{l}h).
\end{align}
Hence, we have
\begin{align}
    &E_{0}\left[\rho^{(1)}( \mathbf{h} )\right]
    =E_{0}\left[\min \limits_{l}\sigma_{l}(c_{l}h)\right]
    =\sum_{t=0}^{\infty}{\prod_{i=1}^{L}{P_{0}(\sigma _{i}(c_{i}h)>t)}} \nonumber\\
    &\geq \sum_{t=0}^{\infty}{P_{0}(\sigma_{1}(c_{1}h)>t)} \nonumber\\
    &-\sum_{t=0}^{\infty}{\sum_{j=2}^{L}{P_{0}(\sigma_{j}(c_{j}h)\leq t <\sigma_{1}(c_{1}h))\prod_{i\neq 1,j}{P_{0}(\sigma _{i}(c_{i}h)>t)} } }. \label{bdd_sigma}
\end{align}
However, the second term of \eqref{bdd_sigma} will vanish as $h \rightarrow \infty$ since from \eqref{ADD}, as $h \rightarrow \infty$,
\begin{align}
P_{0}\left(\sigma_{l}(c_{l}h)\sim\frac{c_{l}h}{\mathcal{I}_{l}}\right)= 1, \label{ADD_CCUSUM}
\end{align}
and with the property of almost sure convergence \cite[Lemma 29, p. 300]{ref:Watanabe18_prop_as}, it can be deduced that  for all $2\leq j \leq L$,
\begin{align}
P_{0}\left(
\sigma_{1}(c_{1}h)-\sigma_{j}(c_{j}h) \sim
\left( \frac{c_{1}}{\mathcal{I}_{1}}-\frac{c_{j}}{\mathcal{I}_{j}} \right)h
<0\right) = 1.
\label{eq_as_sigma}
\end{align}
Combining \eqref{bdd_sigma} and \eqref{eq_as_sigma} then proves \eqref{LADD_malarm}. We then turn to the proof of \eqref{GADD_mvote}. Examining the relationship of CUSUM statistics and log-likelihood ratio shows that
\begin{align}
\rho_N( \mathbf{h} ) = \inf\{t: W_{t}^{k,l}>c_{l}h \textrm{ for all } (k,l)\} \nonumber\\
\leq
\inf\{t: Z_{t}^{k,l}>c_{l}h \textrm{ for all } (k,l)\}. \label{bdd_LLR}
\end{align}
The upper bound in \eqref{GADD_mvote} can then be established by \cite[Lemma 33]{ref:Fellouris18}.
\end{IEEEproof}

\subsection{Proof of Theorem~\ref{2nd_order}}
\begin{IEEEproof}\label{pf:2nd_order}
We first prove \eqref{2nd_ADD_malarm}. Following \cite{ref:Banerjee16}, we decompose the local CUSUM statistic $W_{t}^{k,l}$ as
\begin{align}
    W_{t}^{k,l} = \sum_{s=1}^{t}{Z_{s}^{k,l}}-\min_{0\leq s \leq t}{\sum_{v=1}^{s}{{Z_{s}^{k,l}}}}.
\end{align}
It has been recognized in \cite{ref:Banerjee16} that under $P_0$, $W_{t}^{k,l}$ is a perturbed random walk since the first term is a random walk with positive drift $\mathcal{I}_{l}$ and variance $\sigma_{l}^{2}$, while the second term converges almost surely to a finite random variable as $t \rightarrow \infty$.
Define the stopping time for the $k$-th sensor of group $l$ with threshold $h_{l}$ by
\begin{align}
    \rho^{k,l}(h_{l}) = \textrm{inf} \{t: W_{t}^{k,l}>h_{l} \}. \label{eq:def_stoptimelocal}
\end{align}
Then, from non-linear renewal theory \cite{ref:Woodroofe82}, for $h_{l}= \mathcal{I}_{l}h$ and as $h \rightarrow \infty$,
\begin{align}
    \frac{\rho^{k,l}(\mathcal{I}_{l}h)-h}{\sqrt{\mathcal{I}_{l}h}} \rightarrow \sqrt{\sigma_{l}^{2}/\mathcal{I}_{l}^{3}}\hat{G}_{k,l}, \label{2nd_asym}
\end{align}
in which $\hat{G}_{1,1} \cdots \hat{G}_{N_{L},L}$ are independent standard Gaussian random variables. Let $G_{k,l} = (\sigma_{l}/\mathcal{I}_{l})\hat{G}_{k,l}$ for all $k$, $l$. Then, \eqref{2nd_asym} becomes
\begin{align}
    \frac{\rho^{k,l}(\mathcal{I}_{l}h)-h}{\sqrt{h}} \rightarrow G_{k,l}. \label{2nd_asym_2}
\end{align}
It follows from the continuous mapping theorem \cite[Lemma 28, p. 300]{ref:Watanabe18_prop_as} that
\begin{align}
    \frac{\rho^{(M)}(\mathbf{h})-h}{\sqrt{h}} \rightarrow G^{(M)}, \label{2nd_asym_3}
\end{align}
where $G^{(M)}$ is the $M$-th order statistics of $G_{k,l}$. The only thing left is to check that $\{\rho^{(M)}(\mathbf{h})-h\}/{\sqrt{h}}$ is uniformly integrable under $P_{0}$ so that  
\begin{align}
    E_{0} \left[ \rho^{(M)}(\mathbf{h})-h  \right] = \xi_{M}\sqrt{h}(1+o(1)), 
\end{align}
which can be readily proved by following the steps in \cite{ref:Banerjee16}. 

We go on to prove \eqref{2nd_ADD_mvote}. We denote by $L^{k,l}(h_{l})$ the last time that the local CUSUM of the sensor ${(k,l)}$ is lower than the threshold $h_{l}$ after the stopping time $\rho^{k,l}(h_{l})$, i.e., $L^{k,l}(h_{l}) = \sup\{t>\rho^{k,l}(h_{l}): W_{t}^{k,l}<h_{l}\}$. It follows that
\begin{align}
    \rho_{M}(\mathbf{h})-\rho^{(M)}(\mathbf{h}) \leq \sum_{(k,l)}{L^{k,l}(h_{l})}.
\end{align}
However, since $Z_{t}^{k,l} \leq W_{t}^{k,l}$ for every $t$ and $(k,l)$, we have $L^{k,l}(h_{l}) \leq \sup\{t>\rho^{k,l}(h_{l}): Z_{t}^{k,l}<h_{l}\}$. With the second moment assumption in \eqref{2nd_moment_assumption}, by following the similar steps in \cite{ref:Banerjee16} it can be shown that 
\begin{align*}
E_{0}[\rho_{M}(\mathbf{h})]-E_{0}[\rho^{(M)}(\mathbf{h})] \leq O(1),    
\end{align*}
which together with \eqref{2nd_ADD_malarm} proves the result.

\end{IEEEproof}

\end{document}